# General formula to design a freeform singlet free of spherical aberration and astigmatism : Comment

J. C. Valencia-Estrada[1*] and J. García-Márquez [2]

[1]*Oledcomm SAS, 10-12, avenue de l'Europe, 78140, Vélizy-Villacoublay, France*
[2]*Laboratoire National de Métrologie et d'Essais, LNE, 29, rue Roger Hennequin, 78190, Trappes, France*
*\*camilo.valencia@oledcomm.net*

**Abstract:** In their article [*Appl. Opt.* **58**, 1010–1015 (2019)] González-Acuña *et al* claimed: "an analytical closed-form formula for the design of freeform lenses free of spherical aberration and astigmatism." However, as we show here, their formula can only be applied when the object and image are both real and the image is inversed, additionally, the refractive index in the object and image media is the same. Here, we present the complete solution of this particular formula.

The formula presented by González-Acuña et al. [1] is used to calculate the second surface of a lens, capable to correct the image from spherical aberrations due to the first lens' surface. We have found that this general formula only works when the object distance $f_a$ <0 and the image distance $f_b$ >0 are both real. Nevertheless, their solution fails when one of the combined planes is virtual, *i.e.*, when $f_a$>0 or $f_b$<0 as the required sign rules are not included in the normal $\hat{\mathbf{n}}_a$ formula (Eq. 2). This formula is required to obtain $\hat{\mathbf{v}}_2$ and its correct direction for any virtual-object's point. A different case appears when the image point is virtual. Virtual images are common in car's front lights and lightings. In some thick lenses with positive magnification the rays can cross internally and Eq. (7) does not include the required signs rules. They have shown examples consisting exclusively of real objects and real images.

The authors declare that: "the normal is perpendicular to the tangent plane of the input surface at the origin." and that: "We recall that a necessary condition for the validity of Eq. (7) is that the surface normal should be perpendicular to the tangent plane to the input surface at the origin." This is the definition of the normal and not one condition. However, the condition is that the normal at the origin must be on the optical axis $\hat{\mathbf{n}}_a\big|_{(0,0)} = \pm[0,0,-1]$, which restricts its use to a family of freeform surfaces. This is contradictory with the phrase on the arbitrarily of the surface freeform: "These equations may look cumbersome, but it is quite remarkable that they could be expressed in closed form for an arbitrary freeform input surface."

The fact that "the validity of Eq. (7) also requires that the rays do not intersect each other inside the lens because, in this case, $\Psi_b$ overlaps itself, leaving from being homeomorphic with respect to $\Psi_a$, and the vicinity of the neighborhoods is not preserved" shows that the solution is not complete and that it can only be used to design partially freeform lenses. As a matter of fact, as the sign's rule has not been taken into account, Eq. (7) only applies when the rays do not cross in the interior of the lens. Nevertheless, there are two different families of solution, one having a negative magnification and the second with positive magnification.

These solutions could coexist when the fan of rays is internally split. Selection of the stop aperture position needs to be established for permitting rays from a single family of solutions to enter into the optical system.

In Fig. 1 we can see an example of a freeform lens having internal rays' inversion. Rays cross in its interior and an image free of spherical aberrations is formed.

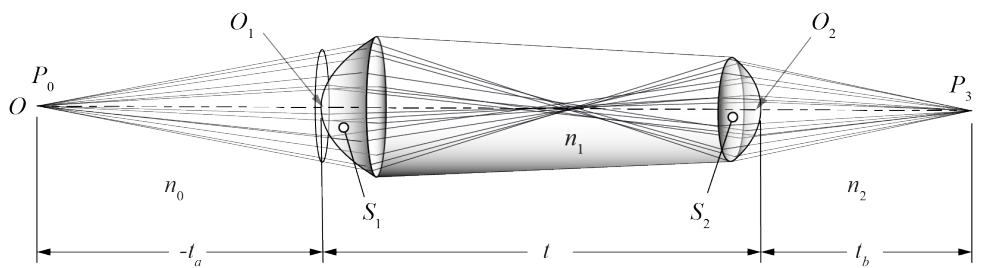

Fig. 1. A lens presenting an internal fan of rays crossing internally. The lens is free of any spherical aberration order and has a positive magnification. From ref [2].

González-Acuña states: "From a more mathematical point of view, since the freeform lens is a homogeneous optical element, the input and output surfaces are simple connected sets on $\mathbb{R}^3$ that can be defined as Eq. (9) are homeomorphic, which means that both surfaces are topologically equivalent. Thus, there exists a continuous and bijective function $f$ such that $f$: $\Psi_a \to \Psi_b$, and whose inverse $f^{-1}$ is also continuous." However, in our opinion this definition is not correct, as several of freeform lenses do not satisfy this definition. The most common case is an aplanatic lens having an image in a single point on the second surface.

The general solution in Eq. (7) fails when the directional cosine $\mathcal{Z} = 0$, *i.e.*, for internal rays traveling on a plane that is perpendicular to the optical axis. Therefore, the transformation is not bijective as the correspondence is neither one-to-one, nor surjective.

The authors' statement on indirect measures of spherical aberration: "We compute the efficiency for 500 rays for all the examples presented in the paper, and the average of all the examples is 99.99999999999410% ≈ 100%". Notwithstanding, all the examples carried out by the authors would undoubtedly correctly perform because they were carefully chosen. In figure 2 (b), a single example of a lens were Gonzalez-Acuña's method fail is shown.

On the authors' opinion about the fact that general Eq. (7) allows for designing singlet freeform lens free of spherical aberration and astigmatism, we think that the question to ask is: is there a spherical aberration-free lens with an astigmatic image on the optical axis? We believe that it is redundant to conclude that the lens is free of spherical aberration and astigmatism on the optical axis. They state that: "it is important to mention that since Eq. (7) in general is not radially symmetric and is not a circle in the $x$–$y$, it also eliminates the astigmatism." However, in a surface or wavefront point, astigmatism is inexistent if the main curvatures are equal.

The solution presented by González-Acuña *et. al.* does not satisfy the fact that: "The formula gives the exact analytical equation of the output surface, given the arbitrary freeform expression of the input surface to correct the spherical aberration and the astigmatism introduced by the first surface." González-Acuña's solution is not concise and the authors do not offer any recommendation to convert the extended parametric solution into a succinct explicit solution. They ignore field aberrations that can grow considerably as it is the case of coma and astigmatism. On the difficulty of maintaining spherical aberration below the diffraction limit in a rotationally symmetric lens, we refer to [4-6].

The one-axis freefrom partial solution of Gonzalez-Acuña *et al* [1] is absolutely referenced as its origin is at the vertex of the anterior surface. It is valid exclusively for

$\hat{\mathbf{n}}_a\big|_{(0,0)} = \pm[0,0,-1]$. Let us follow the notation and conventions of Gonzalez-Acuña *et al* [1] for some variables. It is described in the Table 1 in the Appendix A.

The complete solution for this particular family of one-axis freeform lenses is obtained by applying the method described in [2]. Note that in equation (7) in reference [2] a sign function that ensures the unit vector follows the direction of the ray is embedded, $\text{Sign}(\mathbf{r}_0 \cdot \mathbf{r}_1) = -f_a / (f_a^2)^{1/2}$, $\forall f_a | f_a \neq 0$; $\mathbf{a}_0 = [x_a, y_a, z_a - f_a] / (x_a^2 + y_a^2 + (z_a - f_a)^2)^{1/2}$. Thus, the incident unit vector for this particular case becomes

$$\hat{\mathbf{v}}_1 = -\frac{f_a[x_a, y_a, z_a - f_a]}{\sqrt{f_a^2(x_a^2 + y_a^2 + (z_a - f_a)^2)}} \quad . \tag{1}$$

By following the method described in reference [2], its equation (7) corresponds to equation (1) in this comment. It is interesting to point out that Eq. (1) also corresponds to the variable **s** in reference [3]. The above formula permits one to switch unit vectors direction when the object point is either real or virtual.

The unit vector $\hat{\mathbf{n}}_a$ normal to the first interface is

$$\hat{\mathbf{n}}_a = \frac{[z_{a_x}, z_{a_y}, -1]}{\sqrt{z_{a_x}^2 + z_{a_y}^2 + 1}} \quad , \tag{2}$$

if and only if $z_a = f(x_a, y_a)$. Note that Eq. (2) corresponds to the variable $-\mathbf{g}$ in reference [3].

By applying the vector form of Snell-Descartes' law using dot products in the arbitrary incidence point

$$\hat{\mathbf{v}}_2 = \frac{n_a}{n_l}\left(\hat{\mathbf{v}}_1 - (\hat{\mathbf{n}}_a \bullet \hat{\mathbf{v}}_1)\hat{\mathbf{n}}_a\right) - \left(\sqrt{1 - \frac{n_a^2}{n_l^2}\left(1 - (\hat{\mathbf{n}}_a \bullet \hat{\mathbf{v}}_1)^2\right)}\right)\hat{\mathbf{n}}_a \quad . \tag{3}$$

This formula has the following advantages: a) it is numerically faster, and b) it is valid also for lenses having meridional representation. It is interesting to point out that Eq. (3) also corresponds to the variable $\mathbf{s}''$ in reference [3].

The vector solution is represented with a position vector $\hat{\mathbf{p}}_b$ describing the second surface's geometry of the lens.

$$\hat{\mathbf{p}}_b = [x_b, y_b, z_b] = [x_a, y_a, z_a] + \left(\frac{G'}{V' - \sqrt{V'^2 + \frac{n_l}{n_b}\left(\frac{n_l^2}{n_b^2} - 1\right)f_b G'}}\right)\hat{\mathbf{v}}_2 . \tag{4}$$

with the recurrent variables

$$\begin{cases} \hat{\mathbf{a}}_3 = [-x_a, -y_a, T+f_b-z_a] \\ A' = \dfrac{n_a}{n_b}\left(\sqrt{f_a^2(x_a^2+y_a^2+(z_a-f_a)^2)}/f_a - f_a\right) + \dfrac{n_l}{n_b}T + f_b, \\ V' = \dfrac{n_l}{n_b} f_b\left(\hat{\mathbf{v}}_2 \bullet \hat{\mathbf{a}}_3 - \dfrac{n_l}{n_b}A'\right), \\ G' = \dfrac{n_l}{n_a} f_b\left(\hat{\mathbf{a}}_3 \bullet \hat{\mathbf{a}}_3 - A'^2\right). \end{cases} \qquad (5)$$

Equation (4) in this comment article was deduced from Eq. (19) and (21) in reference [2] by taking into consideration the following:

a) In Gonzalez-Acuña's method, the lens is located in the space with origin at the vertex of the anterior surface. The stop aperture plane $(x_a, y_a, 0)$ is tangent at the same point.
b) In Gonzalez-Acuña's method, the lens is immersed in the same refractive index. Thus, $n_0=1$, $n_1=n$ and $n_2=1$. However, We have the lens immersed in differents media. Thus, $n_0=n_a$, $n_1=n_l$ and $n_2=n_b$.
c) Sign $s_0$ depends on the object position, direction and magnification. For this comment article, this sign corresponds to $s_0 = -(f_a^2)^{1/2}/f_a, \forall \{f_a\}|f_a \neq 0$.
d) Sign $s_2$ depends on the image position, direction and magnification. In this comment article, this sign corresponds to $s_2 = (f_b^2)^{1/2}/f_b, \forall \{f_b\}|f_b \neq 0$.
e) Signs $s_3$ depends on the image position, direction and magnification. For this comment article, this sign corresponds to $s_3 = ((f_b^2)^{1/2}/f_b)((n_b(n_l/n_b)^2)^{1/2})/n_l\}, \forall \{f_b\}| f_b \neq 0$.
f) The recurrent variables $A$, $G$, and $V$ of Eq. (21) in reference [2] were evaluated with the signs $s_0$ and $s_2$. Its results and the sign $s_3$ were substituted in Eq. (19) in reference [2]. The new recurrent variables $A'$, $G'$, and $V'$ were obtained with a process of simplification.

The set of equations (1-5) are completely free of rules' signs. This solution is also advantageously completely differentiable with respect to conjugated distances. Equation (5) is valid for any refractive material. It is valid for any object point, regardless it is real or virtual; it is also valid for any real or virtual image point as well as for any combination of conjugated points, provided the condition be fulfilled,

$$\left| -\dfrac{\partial \hat{\mathbf{p}}_b}{\partial x_a} \times \dfrac{\partial \hat{\mathbf{p}}_b}{\partial y_a} \right| \neq [0,0,0] \quad \forall (x_a, y_a) \subset \text{Aperture}, \qquad (6)$$

which guarantees avoidance of discontinuities and auto-intersections. It is valid for each aperture sector in lenticular anterior surfaces –i.e. in Fresnel surfaces. The set of equations presented here are valid for rays crossing internally; indeed, this permit to design lenses with positive magnification.

A numerical example of the calculation of the aspherical lens with virtual conjugated planes is shown in Fig (2). In 2a) the solution found with the formulation exposed in this comment article is depicted, and 2b) shows the result Gonzalez-Acuña's formulation provides. It is evident that Gonzalez-Acuña's method fails and that the failure origin is the missing sign rules is its formulation. In this example we have chosen as anterior surface an aspero-cylindrical one mathematically represented as

$$z_a(x_a, y_a) = \frac{c_x x_a^2 + c_y y_a^2}{1 + \sqrt{1 - (1+K_x)c_x^2 x_a^2 - (1+K_y)c_y^2 y_a^2}}, \tag{7}$$

where $c_x$ = –1/100 [mm$^{-1}$], and $c_x$ = 1/100 [mm$^{-1}$] are the meridional curvatures in the *XZ* plane and in the plane *YZ* respectively; $K_x$ = 0.5 and $K_y$ = –1.4 are conic constants. The refractive indices are $n_a$ = 1, $n_l$ = 1.5 and $n_b$ = 1, $f_a$= 80 mm is the virtual object distance, $T$ = 10 mm is the lens' central thickness, $f_b$= –50 mm is the virtual image distance, and $d_l$ = 30 mm is the lens' diameter. The stop aperture diameter was set as 25 mm and the number of rays is 21. A comparative example of the calculation of this bi-aspherical cylindrical lens with virtual conjugated planes is shown in Fig (2). The example shows that when applying Eqs. (1-6) the resulting corrective surface works correctly as it is clearly depicted in Fig. 2(a). Nevertheless, it fails when the formula of Gonzalez-Acuña is applied as shown in Fig. 2(b). We can clearly see that the corrective back surface calculated with Gonzalez-Acuña's formulation intercepts the optical axis in the wrong side, *i.e.* before the anterior surface.

It is important to point out that the set of equations (1-5) can be applied for designing one-axis bi-aspheric, rotationally symmetric and spherical aberration-free lenses. Particular solutions in [5-7] are not robust nor are complete nor closed [8]. Table 2 in the Appendix B lists the variables and an operator used in references [5-7]. By substituting $\{x_a \rightarrow r_a \cos \theta\}$ and $\{y_a \rightarrow r_a \sin \theta\}$ in Eqs. (1-5) it is possible to obtain a set of parametric equations in cylindrical coordinates. Given that the surface is rotationally symmetric, parameter $\theta$ vanish in the simplification process by using the fact that $\cos^2 \theta + \sin^2 \theta = 1$. In this manner, we obtain a meridional representation ($r_a$, $z_a$) with the variable change $\{x_a^2 + r_a^2 \rightarrow r_a^2\}$. Thus, we can obtain a more robust, complete and rigorous formula to design one-axis bi-aspheric lenses immersed in a single refractive index:

a) A negative lens calculated by using this article's formulation

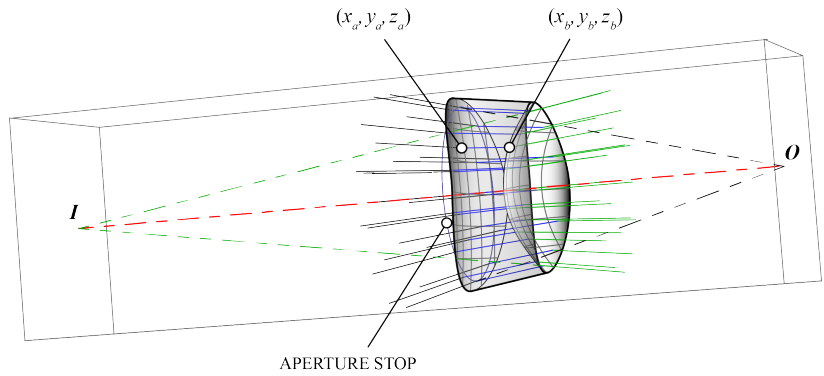

b) The surface found through Gonzalez-Acuña's formulation is not a solution

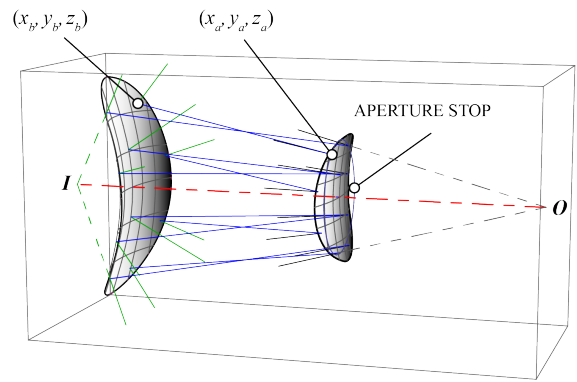

Fig. 2. A comparative example. Extended phantom lines (rays) indicate the virtual object and image conjugated planes. In (b) rays are generated in wrong directions.

$$\begin{cases}
\boldsymbol{v}_1 = -\dfrac{t_a[r_a, z_a - t_a]}{\sqrt{t_a^2(r_a^2 + (z_a - t_a)^2)}}, \\[6pt]
\boldsymbol{n}_a = \dfrac{[z'_a, -1]}{\sqrt{z'^2_a + 1}}, \\[6pt]
\boldsymbol{v}_2 = \dfrac{n_a}{n_l}\left(\boldsymbol{v}_1 - (\boldsymbol{n}_a \bullet \boldsymbol{v}_1)\hat{\boldsymbol{n}}_a\right) - \left(\sqrt{1 - \dfrac{n_a^2}{n_l^2}\left(1 - (\boldsymbol{n}_a \bullet \boldsymbol{v}_1)^2\right)}\right)\boldsymbol{n}_a, \\[6pt]
\boldsymbol{a}_3 = [-r_a, t + t_b - z_a], \\[6pt]
A' = \dfrac{n_a}{n_b}\left(\sqrt{t_a^2(r_a^2 + (z_a - t_a)^2)}/t_a - t_a\right) + \dfrac{n_l}{n_b}t + t_b, \\[6pt]
V' = \dfrac{n_l}{n_b}t_b\left(\boldsymbol{v}_2 \bullet \boldsymbol{a}_3 - \dfrac{n_l}{n_b}A'\right), \\[6pt]
G' = \dfrac{n_l}{n_b}t_b\left(\boldsymbol{a}_3 \bullet \boldsymbol{a}_3 - A'^2\right), \\[6pt]
\boldsymbol{p}_b = [r_b, z_b] = [r_a, z_a] + \left(\dfrac{G'}{V' - \sqrt{V'^2 + \dfrac{n_l}{n_b}\left(\dfrac{n_l^2}{n_b^2} - 1\right)t_b G'}}\right)\boldsymbol{v}_2.
\end{cases} \quad (8)$$

The formula (8) is valid for any object point | $t_a \neq 0$, regardless it is real or virtual; it is also valid for any real or virtual image point | $t_b \neq 0$; in other words, it is valid for any combination of conjugated points with non-null distances. The formula (8) is valid if back surface normals exist $\forall r_a \subset$ Aperture as we have shown in [2].

The code (found in the supplementary material) typed for Mathematica® is used to design one-axis freeform lenses free of spherical aberrations using this article's Eqs. (1-5). This code can also be applied to design bi-aspherical lenses.

# APPENDIX A

**Table 1. Variables and Operator's Descriptions**

| Variables and operator | Description |
|---|---|
| $x_a$ | Abscissa of the anterior surface in Cartesian coordinates. |
| $y_a$ | Abscissa of the anterior surface in Cartesian coordinates. |
| $z_a$ | Ordinate of the anterior explicit surface in Cartesian coordinates. |
| $x_b$ | Abscissa of the back surface in Cartesian coordinates with parameters $x_a$ and $y_a$. |
| $y_b$ | Abscissa of the back surface in Cartesian coordinates with parameters $x_a$ and $y_a$. |
| $z_b$ | Axial coordinate of the back surface in Cartesian coordinates with parameters $x_a$ and $y_a$. |
| $\hat{\mathbf{v}}_1$ | Unit vector of the incident ray. |
| $\hat{\mathbf{v}}_2$ | Unit vector of the refracted ray. |
| $\hat{\mathbf{n}}_a$ | Normal unit vector to the first surface. |
| $z_{a_x}$ | Partial derivative of the ordinate of the anterior surface with respect to $x_a$. |
| $z_{a_y}$ | Partial derivative of the ordinate of the anterior surface with respect to $y_a$. |
| $f_a$ | Object distance (following Descartes' rule of signs). |
| $T$ | Center thickness. |
| $f_b$ | Image distance (following Descartes' rule of signs). |
| $n$ | Relative refractive index of the lens with respect to external media. |
| $\hat{\mathbf{a}}_3$ | Vector between an arbitrary point of the anterior surface and the image point. |
| $A', V', G'$ | Recurrent parametrical variables. |
| $\hat{\mathbf{p}}_b$ | Position vector of the back surface in parametrical coordinates. |
| $\bullet$ | Dot product of vectors. |

# APPENDIX B

**Table 2. Variables and Operator's Descriptions**

| Variables and operator | Description |
|---|---|
| $r_a$ | Radial abscissa of the anterior surface in cylindrical coordinates. |
| $z_a$ | Ordinate of the anterior explicit surface in cylindrical coordinates. |
| $r_b$ | Radial coordinate of the back surface in cylindrical coordinates with parameter $r_a$. |
| $z_b$ | Axial coordinate of the back surface in cylindrical coordinates with parameter $r_a$. |
| $\boldsymbol{v}_1$ | Unit vector of the incident ray. |
| $\boldsymbol{v}_2$ | Unit vector of the refracted ray. |
| $\boldsymbol{n}_a$ | Normal unit vector to the first surface. |
| $z'_a$ | Derivative of the ordinate of the anterior surface with respect to $r_a$. |
| $t_a$ | Object distance (following Descartes's rule of signs). |
| $t$ | Center thickness |
| $t_b$ | Image distance (following Descartes's rule of signs). |
| $n$ | Relative refractive index of the lens with respect to external media. |
| $\boldsymbol{a}_3$ | Vector between an arbitrary point of the anterior surface and the image point. |
| $A', V', G'$ | Recurrent parametrical variables. |
| $\boldsymbol{p}_b$ | Position vector of the back surface in parametrical coordinates. |
| $\bullet$ | Dot product of vectors. |

**Supplementary material: Code for Mathematica® :**

(*         **One-axis freefrom and bi-aspheric lenses**         *)

(*               **without spherical aberration**               *)

(*              **Code developed for Mathematica by**           *)

(*        **J.C. Valencia-Estrada[1] and J. Garcia-Marquez[2]** *)

(*

[1] Oledcomm SAS,10-12, avenue de l'Europe, 78140,Vélizy-Villacoublay, France

   camilo.valencia@oledcomm.net

   juanvalenciaestrada@gmail.com

[2] Laboratoire National de Métrologie et d'Essais (LNE), 29 rue Roger Hennequin, 78190, Trappes, France

*)

**Clear**["Global`*"]

offset = 20;

(* Input variables *)

ta = 40;                          (* Object distance *);

t = 80;                           (* Central thickness *);

tb = 50;                          (* Image distance *);

fa = ta;

fb = tb;

T = t;

na = 1.333;                       (* Anterior refractive index *);

nl = 1.724;                       (* Lens's refractive index *);

nb = 1;                           (* Back refractive index *);

d = 30;                           (* Anterior surface diameter *);

p[0] = {0, 0, ta};                (* Object point *);

o[1] = {0, 0, 0};                 (* Anterior surface vertex point *);

**(*Anterior surface's vector position *)**

p[1] ={x, y, (cx x^2 + cy y^2) / (1 + **Sqrt**[1- (1 + Kx) cx^2 x^2 - (1 + Ky) cy^2 y^2])};

**(* Anterior surface's parameters *)**

Kx = 0.5; Ky = -1.4;                      (* Meridional conical constants *);

cx = 1/100; cy = 1/20;                    (* Meridional vertex curvatures *);

**(* Anterior surface with explict representation *)**

za = (cx x^2 + cy y^2) / (1 + **Sqrt**[1 - (1+ Kx) cx^2 x^2 - (1 + Ky) cy^2 y^2]);

s[1] = {0, 0, 0}; dstop = 25; rstop = dstop/2;      (* Stop plane position and diameter *);

o[2] = {0, 0, t};                         (* Back surface vertex point *);

p[3] = {0, 0, t + tb};                    (* Image point *);

**(* Path vectors: For more information please follow [2] *)**

r[0] = {0, 0, -ta};

r[1] = {0, 0, t};

r[2] = {0, 0, tb};

a[0] = p[1] - p[0];

a[4] = p[3] - p[1];

**Print**["Anterior surface      = ", p[1]]

**Print**["Domain                = ", domain = x^2 + y^2 < d^2/4]

**(* Pupil or stop routine for 11 rays *)**

tabla =

{

{-0.262258924190165855095630653709, -0.689552138434555425611558523406},

{0.262258924190165855095630653709, -0.689552138434555425611558523406},

{-0.654207495490543857031888931623,-0.340990392505480262378855842125},

{0.654207495490543857031888931623, -0.340990392505480262378855842125},

{0, -0.235306356998833687832605108517},

{-0.715460686241806569843043724712, 0.179938604472344027862805139398O},

{0.715460686241806569843043724712, 0.179938604472344027862805139398O`},

{0, 0.289211491381498022358656l989},

{-0.415055617900124834285924684739, 0.609910427019080420778753583334},

{0.415055617900124834285924684739, 0.609910427019080420778753583334},

{0,0}

};

(* Scaling *)

rmax = 0;

Do[rmesur = Sqrt[ tabla[[j, 1]]^2 + tabla[[j, 2]]^2 ];

If [rmesur > rmax, rmax = rmesur], {j, 1, 11} ];

tabla = tabla (rstop/rmax);

stop = {rstop Cos[$\alpha$], rstop Sin[$\alpha$], 0};

b[0] = s[1];

tablar ={};

Do[tablar = Append [tablar, {tabla[[j, 1]], tabla[[j, 2]], 0} ], {j, 1, 11} ];

(* Vector calculations *)

vh1 = - ((fa a[0]) / Sqrt[fa^2 (a[0] . a[0])]);          (* Incident unit vector *);

puntos = {};

Do[vss = tablar[[w]] - p[0]; vs = vss / Sqrt [vss . vss]; tempo = Chop[FindRoot [{vs[[1]] - vh1[[1]] == 0, vs[[2]] - vh1[[2]] == 0}, {x, 0}, {y, 0}]]; ray[w]={x, y}/.{x -> tempo[[1]], y -> tempo[[2]]}; p1[w] = p[1] /.ray[w]; puntos = Append[puntos, p1[w]], {w, 1, 11}];

Do[v0[w] = vh1/.ray[w], {w, 1, 11}];

Do[cp1[w] = Cross[D[p[1], x]/.ray[w], D[p[1], y]/.ray[w]], {w, 1, 11}];

nva = {D[za, x], D[za, y], -1} / Sqrt[{D[za, x], D[za, y], -1} . {D[za, x], D[za, y], -1}];

(* Unit normal vector of the anterior surface *)

Do[n1[w] = {D[za, x]/.ray[w], D[za, y]/.ray[w], -1} / (Sqrt[{D[za, x]/.ray[w], D[za, y]/.ray[w], -1} . {D[za, x]/.ray[w], D[za, y]/.ray[w], -1})], {w, 1, 11}];

vh2 = Chop[na/nl (vh1- (nva . vh1) nva) - nva Sqrt[1 - na^2/nl^2 (1 - (nva . vh1)^2)]] ;

(* Refracted unit vector *)

Do[v1[w] = na/nl (v0[w] - (n1[w] . v0[w]) n1[w]) - n1[w] Sqrt[1- na^2/nl^2 (1 - (n1[w] . v0[w])^2)], {w, 1, 11}];

**(* Recurrent variables *)**

ah3 = p[3] - p[1];

Ah = **Chop**[ na/nb (**Sqrt**[fa^2 (a[0] **.** a[0])] / fa - fa) + nl/nb T + fb];

Vh = nl/nb fb (**Chop**[**Dot**[vh2, ah3] - nl/nb Ah]);

Gh = nl/nb fb (**Chop**[**Dot**[ah3, ah3]] - Ah^2);

**Do**[k1[w] = **N**[na/nb (**Sqrt**[fa^2((p1[w] - p[0]) **.** (p1[w] - p[0]))]/fa - fa) + nl/nb T + fb], {w, 1, 11}];

**Do**[k2[w] = **Dot**[(p[3] - p1[w]), (p[3] - p1[w])], {w, 1, 11}];

**Do**[k3[w] = nl/nb fb (**Dot**[v1[w], (p[3] - p1[w])] - nl/nb k1[w]), {w, 1, 11}];

**Do**[Gar[w] = nl/nb fb (k2[w] - k1[w]^2), {w, 1, 11}];

**(* Back surface's vector position *)**

pa = p[1];

pb = **Chop**[pa + (Gh vh2)/(Vh - **Sqrt**[Vh^2 + nl/nb fb Gh ((nl/nb)^2 - 1)])];

**Print**["Test                = ", o[2] -> **Chop**[pb/.{x -> 0, y -> 0}]]

**Do**[p2[w] = **Chop**[p1[w] + (Gar[w] v1[w]) / ( k3[w] - **Sqrt**[k3[w]^2 + nl/nb fb Gar[w] ((nl/nb)^2 - 1)]) ], {w, 1, 11}];

Dpbx = **D**[pb, x]; Dpby = **D**[pb, y];

**Do**[Dpx2[w] = Dpbx/.ray[w]; Dpy2[w] = Dpby/.ray[w], {w, 1, 11}];

**Do**[N2[w] = **Cross**[Dpx2[w], Dpy2[w]]; n2[w] = -N2[w] / **Sqrt**[N2[w] **.** N2[w]], {w, 1, 11}];

**Do**[v2[w] = **Chop**[nl/nb (v1[w] - (n2[w] **.** v1[w]) n2[w]) - n2[w] **Sqrt**[1- (nl/nb)^2 (1 - (n2[w] **.** v1[w])^2)]], {w, 1, 11}];

**(* Fan 0 *)**

fan[0] = {};

**If**[((**Sign**[r[1] **.** r[0]] == 1) && (nl/na != -1)), **Do**[rayo[w] = **Graphics3D**[{Gray, **Line**[{p[0], p1[w]}], Boxed -> False}]; fan[0] = **Append**[fan[0], rayo[w]], {w, 1, 11}],

lenobj ={};

**Do**[lenobj = **Append**[lenobj, (p1[w] - p[0]) **.** (p1[w] - p[0])], {w, 1, 11}];

robj = **Sqrt**[**Max**[lenobj]] + offset;

**Do**[lamb = robj / **Norm**[p1[w] - p[0]]; ps[w] = p[0] + (p1[w] - p[0]) lamb, {w, 1, 11}];

Do[rayo[w] = **Graphics3D**[{Black, **Line**[{ps[w], p1[w]}], Boxed -> False}]; fan[0] = **Append**[fan[0], rayo[w]], {w, 1, 11}]];

**(* Edge *)**

b[0] = {d/2 Cos[α], d/2 Sin[α], 0};

b[1] = **Chop**[p[1]/.{x -> b[0][[1]], y -> b[0][[2]]}];

b[2] = **Chop**[p[2]/.{x -> b[0][[1]], y -> b[0][[2]]}];

umax = **FindMaximum**[b[0][[1, α]][[1]]];

umin = **FindMinimum**[b[0][[1]],α][[1]];

vmax = **FindMaximum**[b[0][[2]],α][[1]];

vmin = **FindMinimum**[b[0][[2]], α][[1]];

**(* Anterior surface plot *)**

surface[1] = **ParametricPlot3D**[**Chop**[pa/.{x -> u, y -> v}], {u, umin, umax}, {v, vmin, vmax}, Mesh -> None, PlotStyle -> Directive[Opacity[.4], Gray], Boxed -> False, RegionFunction -> Function[{x, y, z, u, v}, u^2 + v^2 <= d^2/4]];

**(* First refraction *)**

**(* Fan 1 *)**

fan[1] = {};

**Do**[rayo[w] = **Graphics3D**[{Black, **Line**[{p1[w], p2[w]}], Boxed -> False}]; fan[1] = **Append**[fan[1], rayo[w]], {w, 1, 11}];

**(* Back surface plot *)**

surface[2] = **ParametricPlot3D**[**Chop**[pb/.{x -> u, y -> v}],{u, umin, umax},{v, vmin, vmax}, Mesh -> None, PlotStyle -> Directive[Opacity[.4], Gray], Boxed -> False, RegionFunction -> Function[{x, y, z, u, v}, u^2 + v^2 <= d^2/4]];

**(* 2nd refraction *)**

**(* Fan 2 *)**

fan[2] = {};

**If**[((**Sign**[r[2] . r[1]] == 1) && (nl/nb != -1 )), **Do**[rayo[w] = **Graphics3D**[{Gray, **Line**[{p2[w], p[3]}], Boxed -> False}]; fan[2] = **Append**[fan[2], rayo[w]], {w, 1, 11}],

lenimg2 = {};

**Do**[lenimg2=**Append**[lenimg2, (p2[w] - p[3]) . (p2[w] - p[3])], {w, 1, 11}]; rimg2 = **Sqrt**[**Max**[lenimg2]] + offset;

**Do**[lamc2 = rimg2/**Norm**[p2[w] - p[3]]; pf[w] = p[3] + (p2[w] - p[3]) lamc2, {w, 1, 11}];

**Do**[rayo[w] = **Graphics3D**[{Gray, **Line**[{p2[w], pf[w]}], Boxed -> False}]; fan[0] = **Append**[fan[0], rayo[w]], {w, 1, 11}]];

reference = **Graphics3D**[{Red, Thickness[0.002], **Line**[{{p[0], o[1]}, {o[1], o[2]}, {o[2], p[3]}}]}];

(* Show *)

**Show**[fan[0], surface[1], fan[1], surface[2], fan[2], reference, **Graphics3D**[{**PointSize**[Small], **Point**[tablar]}], **ParametricPlot3D**[{stop + s[1], b[2]}, {α, 0, 2 Pi}], **Graphics3D**[{**PointSize**[Small], **Point**[puntos]}]]

**Label**[end];

Anterior surface = {x, y, (x^2/100 + y^2/20) / (1 + Sqrt [1-0.00015 x^2 + 0.001 y^2])}

Domain = x^2+y^2 < 225

Test = {0,0,80} -> {0,0,80.}

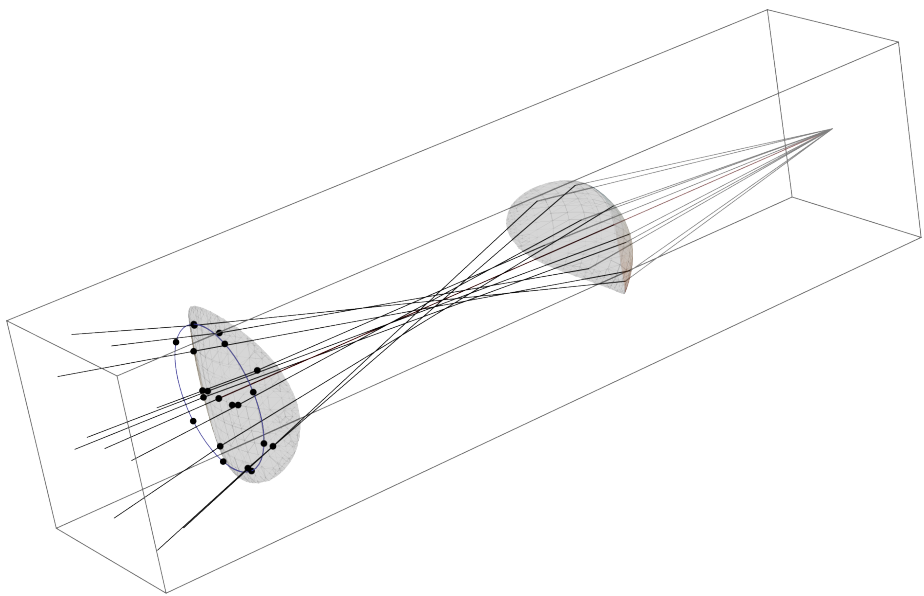